\def\beq{\begin{equation}}

\def\n{\nonumber}
\newif\iffigs\figsfalse
\figstrue

\iffigs

\else

\fi
\def\bbbz{{\bf Z}}

\def\bbb1{{\rm 1\!1}}

\newcommand{\refs}[1]{(\ref{#1})}

 \documentclass[12pt]{JHEP}

\usepackage{epsfig}
\advance\hoffset by -.35in
\def\pa{\partial}
\def\es{\!=\!}
\def\ha{{1\over 2}}
\def\>{\rangle}
\def\<{\langle}
\def\mtx#1{\quad\hbox{{#1}}\quad}

\def\A{{\cal A}}

\def\de{\delta}

\def\F{{\cal F}}
\def\Y{{\cal Y}}

\def\al{\alpha}

\def\Tr{\hbox{Tr}}

\def\la{\lambda}

\def\al{\alpha}

\def\de{\delta}

\def\la{\lambda}

\preprint{KCL-TH-00-35}

\title{\Large\bf NON-ABELIAN BORN-INFELD ACTION AND SOLITONS FOR 
CRITICAL NON-BPS BRANES}

\author{N.D. Lambert$^{a}$ and  I. Sachs$^{b}$\\

\vskip 24pt

$^{a}$Dept. of Mathematics\\
King's College\\
The Strand\\
London, WC2R 2LS\\
England\\

\vskip 24pt

$^{b}$Theoretische Physik\\
Ludwig-Maximilians Universit\"at\\
Theresienstrasse 27 \\80333 Munich\\ Germany }
\abstract{
The non-abelian flat directions in the tachyon potential of  stable 
non-BPS branes recently found are shown to persist to 
all orders in $\al'$ at tree level in the string coupling. 
We also obtain the non-abelian Born-Infeld action 
including the tachyon potential for a stack of stable non-BPS branes
on a critical orbifold. Finally we 
discuss stable soliton states on the non-BPS brane.}

\begin{document}
\section{Introduction}

D-branes play an important role in understanding 
non-perturbative 
aspects of string theory and supersymmetric field theory alike 
\cite{Polchinski94}.
While the dynamics of D-branes 
are encoded in the open strings ending on them, the low 
energy effective 
action, to leading order in $\al'$, has been shown to be that of 
dimensionally reduced, 
ten-dimensional supersymmetric Yang-Mills theory \cite{Witten2}. For constant 
field strength and as long as the second derivatives of the scalars in the 
Yang-Mills mutiplet are small, higher order 
corrections in $\al'$ are  described in terms of the Born-Infeld 
action \cite{Fradkin,BSPT,AT,Tseytlin,Callan,Leigh,Perry,Myers}. 

More recently it has become clear that in addition to D-branes 
the spectrum type II string theory 
contains non-supersymmetric but, nevertheless, 
stable states 
\cite{sen9803194,sen9805019,sen9812031,sen9904207,Bergman9806155,Bergman9901014}. At leading order the dynamics of these non-BPS D-branes 
is again described by a field theory. However, unlike for D-branes, 
this field theory is not obtained from dimensionally reduced ten-dimensional 
Yang-Mills theory. In particular, there are extra scalars which originate 
in a tachyonic sector of the open strings ending on these branes. 
To leading order in $\al'$ the field theory of a stack of stable non-BPS 
D-branes was obtained in \cite{LS}. 
A, maybe unexpected, feature of this theory is that at certain points 
in the moduli 
space of stable non-BPS D-branes the potential for the 
``Tachyons''\footnote{Note that although we refer to certain scalar fields 
of a non-BPS D-brane as tachyons, due to an orbifold projection, these
modes are in fact massless.} has 
non-abelian flat directions. 
An immediate question is, of course, whether or not 
these flat directions, which are not protected by supersymmetry, 
are removed by higher order corrections. On general grounds, we expect that 
$\al'$ corrections to the field theory approximation should 
again be described by some Born-Infeld action. The purpose of 
the present paper therefore is to construct this action, to all orders in 
$\al'$, for a stack of stable non-BPS D-branes. The challenging 
part of this construction is to quantise open strings in a tachyonic 
background. For a generic point of the moduli space this is indeed a 
difficult problem \cite{Fendley,Callan9402113,Thorlacius9404008}. However, 
at the point in moduli space where the tachyons become massless, 
the problem of quantising open strings in a constant tachyon background 
can be mapped to that of a non-abelian Wilson-line background by a suitable 
change of variables \cite{Sen9808141,Sen0003124}. In the new variables the 
condition for vanishing of the potential for the tachyon becomes simply a 
zero curvature condition which we will show to be equivalent to the 
non-abelian flat directions found in \cite{LS}. 

In the analysis of brane dynamics, an important role is played by 
the solitonic excitations on
them. These usually have a spacetime interpretation
as intersecting branes and provide information on some non-perturbative
aspects of the worldvolume theory. These worldvolume solitons are 
also important in applications of D-branes
to the study of Yang-Mills dynamics. Based on the worldvolume effective action 
constructed here we also include
a discussion of the stable solitons on  non-BPS
D-branes. 

Thus in this paper, we will be interested in the tree-level effective 
action for
a stack of stable non-BPS branes. BI actions for unstable D-branes are 
discussed in \cite{Sen9909062,Garousi}. For the techincal reason mentioned 
above we will
restrict our attention to the critical radius. We will also discuss
some features of the stable soliton solutions on the non-BPS brane. 
Indeed it would appear that the more interesting solitons only arise at the
critical radius.  We leave the issue of string
loop corrections and the effective potential to future work \cite{LSprep}.

The plan for the rest of the paper is as follows. In the next section we
briefly review the construction of a non-BPS D-brane in type II string theory.
In section three we then adapt the change of variables in 
\cite{Sen9808141,Sen0003124} to our present case. This then allows us to
derive  the complete low energy effective action in section four, although
we will see that it is inherently non-abelian and so suffers the same
problems that are encountered with the non-abelian Born-Infeld action for 
D-branes \cite{Tseytlin,Perry}. Finally in section five we discuss some of the stable soliton states on a
non-BPS brane and in particular their spacetime interpretation.

\section{Review of Stable Non-BPS Branes Wrapped on $T^{4}/\bbbz_{2}$.}

We start by briefly recalling the key features
non-BPS D$p$-branes as presented in the reviews articles 
\cite{sen9904207,Gaberdiel}. A non-BPS D$p$-brane (${\tilde D}p$-brane) 
can be defined as having
two types of open strings ending on it, labelled by the Chan-Paton
indices $I$ and $\sigma_1$. Contrary to BPS D$p$-branes one has $p$ even
in type IIB string theory and $p$ odd in type IIA string theory.
The $I$-sector strings are precisely the
same as for BPS D-branes and therefore their low energy modes form
a maximally supersymmetric vector mulitplet in $(p+1)$ dimensions.
The $\sigma_1$-sector strings 
come with the opposite projection under $(-1)^F$.
Therefore, after the GSO projection, the level one sector modes of
the NS-sector are projected out while the tachyonic ground state survives.
Thus the lightest modes of the $\sigma_1$-sector 
consist of real tachyonic scalar with mass-squared
$-1/2\alpha'$ and a sixteen-component fermion. 

Clearly this system is unstable due to the tachyon. However by wrapping
the brane around an orbifold\footnote{Note that here we are working in the
T-dual picture to \cite{LS} where the branes were not wrapped around 
the orbifold.}, for example we will take 
$T^{4}/\bbbz_{2}$ with coordinates  $i\es 6,\ldots,9$ and radii $R_i$, 
the tachyonic modes which are even under the orbifold action 
$g:x^i\leftrightarrow -x^i$ are projected out. Thus only the 
tachonic modes with odd momentum around the orbifold survive. Since
a momemtum mode has mass-squared
\begin{equation}
m^2 = -{1\over 2\alpha'}+ \sum_{i=6}^9 \left(n_i\over R_i\right)^2\ ,
\end{equation}
we see that below a critical radius $R_c = \sqrt{2\alpha'}$ the 
surving states are non-tachyonic. 
In particular,
with the conventions of \cite{LS}, the four lightest tachyon scalars surviving 
the orbifold are given by (we set $\alpha'=1$)
\begin{equation}
        T^{i}=\frac{\chi^{i}(x)}{2i}\left(e^{i\vec{\omega}^{i}\cdot
        \vec{X}}-e^{-i\vec{\omega}^{i}\cdot\vec{X}}\right),
        \label{v0}
\end{equation}
where $\vec{\omega}^{i}\es  {(R_i)^{-1}\vec{e}^{i}}$. In this way we
obtain a stable non-BPS brane.

On the other
hand, the effect of the orbifold on the bosons of the $I$-sector strings
are the same as for a D$p$-brane at an orbifold. Therefore, in the
case of a D-sevenbrane wrapped on $T^{4}/\bbbz_{2}$, 
which we will be most interested in here, 
the $N=4$ vector multiplet is reduced to an $N=2$ multiplet. This contains two
scalars $X^I$, $I=4,5$, that represent the fluctuations of the brane
in the $x^4$ and $x^5$ directions, and a vector $A_\mu$, $\mu=0,1,2,3$. 
For the fermions the orbifold selects out  one chirality under
$\Gamma^{6789}$ for the $I$-sector and the opposite for the $\sigma_1$-sector.
This is the same fermionic content as four-dimensional $N=4$  
super-Yang-Mills.
However even at the critical radius where the tachyonic
modes are massless and the field content is precisely that of $N=4$
super-Yang-Mills, the effective theory is not supersymmetric \cite{LS}.

For the rest of this paper we will restrict our attention to oribifolds at 
the critical radius $R_i=\sqrt{2}$ where the lightest tachyon modes 
are massless.  
We will only use the Einstein summation convention
for the full ten-dimensional indices $m,n=0,1,2...,9$ and the
four-dimensional indices $\mu,\nu=0,1,2,3$. 
All other sums (e.g. those over the
$i,j$ and $I,J$ indices) will be explicitly written.

\section{Open Strings in a Tachyon Background}

In this section we adapt the change of variables introduced in 
\cite{Sen9808141,sen9904207,Sen0003124} to the present situation. For 
simplicity we first consider the case of an abelian tachyon (i.e. 
for one $\tilde D3$-brane on $K3$) and the generators of the $U(N)$ 
Lie algebra at the end. 

The objects of interest are the vertex operators for $T^{i}$ in the 
$(0)$ and $(-1)$ pictures. 
With the conventions of \cite{LS} but with $\al'\es g_{s}\es1$ we then 
have 
\begin{eqnarray}
        V_{T}^{(0)i}&=&\frac{1}{\sqrt{2}}(\vec{\omega}^{i}\vec{\psi})
        \left(e^{i\vec{\omega}^{i}\cdot
        \vec{X}}+e^{-i\vec{\omega}^{i}\cdot\vec{X}}\right)
        \n\\
        &=&\frac{1}{2}\psi^{i}\left(e^{i\frac{1}{\sqrt{2}}
        X^{i}}+e^{-i\frac{1}{\sqrt{2}}X^{i}}\right)\ .
\label{v1}
\end{eqnarray}
To continue we map the tachyon vertex operator \refs{v1} into 
a non-abelian Wilson line \cite{Sen9808141,Sen0003124}. 
Here we will only discuss in detail the
$(0)$ picture but the relevant discussion for the  $(-1)$ picture can be
easily constructed by following the same change of variables but starting
with
\begin{equation}
V_T^{(-1)i} = -\frac{i}{\sqrt{2}}e^{-\Phi}
\left(e^{i\vec{\omega}^{i}\cdot\vec{X}}
-e^{-i\vec{\omega}^{i}\cdot\vec{X}}\right)\ ,
\end{equation}
where $\Phi$ is the bosonised superconformal ghost. 
For this we 
first express $V_{T}^{(0)i}$ in terms of the closed string fields 
$X_{R/L}^{i},\psi_{R/L}^{i}$ with Neumann boundary condition
\begin{eqnarray}
        X^{i}_{L}=X^{i}_{R}=\ha X^{i}\ ,\n\\
        \psi^{i}_{L}=\psi^{i}_{R}=\psi^{i}\ ,      
        \label{v2}
\end{eqnarray}
at both ends in the NS sector. In the R sector \refs{v2} applies at
one end and 
\begin{equation}
        \psi^{i}_{R}=-\psi^{i}_{L}\ ,
        \label{v2.1}
\end{equation}
in the other. The vertex operator \refs{v1} can be written 
as
\begin{equation}
        \frac{1}{2}\psi^{i}_{R}
        \left(e^{i\sqrt{2}X_{R}^{i}}+e^{-i\sqrt{2}X_{R}^{i}}
        \right)
        \label{v2.2}
\end{equation}
At the critical orbifold ($R_{i}\es R_{c}\es\sqrt{2}$) we may 
fermionise $e^{i\sqrt{2}X_{R}^{i}}$ as 
\begin{equation}
        e^{i\sqrt{2}X_{R}^{i}}=\frac{1}{\sqrt{2}}(\xi^{i}_{R}+i\eta^{i}_{R})
        \otimes\Gamma^{i}\mtx{and}
        e^{i\sqrt{2}X_{L}^{i}}=\frac{1}{\sqrt{2}}(\xi^{i}_{L}+i\eta^{i}_{L})
        \otimes\Gamma^{i}\ ,
        \label{v3}
\end{equation}
where the cocycles $\Gamma^{i}$ are introduced to restore the 
correct commutation relations with th world-sheet fermions 
\cite{Sen0003124,Banks}. This can be achieved by taking for 
$\Gamma^{i}$ the generators of the $Spin(4)$-Clifford algebra and 
attaching a $\Gamma^{5}$ to the world sheet fermions. We complete the 
transformation by rebosonising as 
\begin{equation}
        \frac{1}{\sqrt{2}}\left(\xi^{i}_{R/L}\pm 
        i\psi^{i}_{R/L}\right)=e^{\pm i\sqrt{2}\tilde X^{i}_{R/L}}
        \otimes \tilde \Gamma^i
        \ , \label{v4}
\end{equation}
and attaching a $\tilde\Gamma^{5}$ to $\eta^{i}_{R/L}$. Here the 
$\tilde\Gamma^i$ form another representation of the $Spin(4)$-Clifford algebra
which commutes with the $\Gamma^i$ representation.
With these 
conventions the fermionic and bosonic currents are then related as 
\begin{equation}
        \eta^{i}_{R}\xi^{i}_{R}=i\sqrt{2}\pa X^{i}_{R}\mtx{and}
        \psi^{i}_{R}\xi^{i}_{R}=i\sqrt{2}\pa \tilde X^{i}_{R}\ .
                \label{v5}
\end{equation}
The boundary conditions for the different fields are determined as 
follows: From \refs{v1}, \refs{v2} and \refs{v3} we have 
\begin{equation}
        \xi^{i}_{L}=\xi^{i}_{R}=:\xi^{i}\mtx{and} 
        \eta^{i}_{L}=\eta^{i}_{R}=:\eta^{i}.
        \label{v6}
\end{equation}
In the NS sector, where $\psi^{i}_{L}\es\psi^{i}_{R}$ on both ends 
of the open string, \refs{v6} implies
\begin{equation}
        \tilde X^{i}_{L}=\tilde X^{i}_{R}=:\ha\tilde X^{i}\ ,
        \label{v7}
\end{equation}
i.e. NN boundary conditions for $\tilde X^{i}$. In the R-sector, 
where $\psi^i_{R}(\pi)\es-\psi^i_{L}(\pi)$, 
\refs{v6} implies in turn 
\begin{equation}
        \tilde X^{i}_{L}(\pi)=-\tilde X^{i}_{R}(\pi)\ ,
        \label{v8}
\end{equation}
i.e. ND boundary conditions for $\tilde X^i$. Using \refs{v5},\refs{v7} and
\refs{v8} we 
may now write
\begin{eqnarray}
        V^{(0)i}_{T}&=&i\pa_{||}\tilde X^{i}\otimes\sigma_{1}
        \otimes\Gamma^{5}\Gamma^{i}\ , \n\\
        V^{(0)i}_{T}&=&i\pa_{\perp}\tilde X^{i}\otimes\sigma_{1}
        \otimes\Gamma^{5}\Gamma^{i}\ ,
        \label{v9}
\end{eqnarray}
in the NS and R sectors respectively. Finally the
vertex operators $V^{(-1)i}$ in the $(-1)$ picture become simply
\begin{equation}
V_T^{(-1)i} = e^{-\Phi}\eta^i\otimes \sigma_1 \otimes \tilde\Gamma^5\Gamma^i\ .
\end{equation} 

To summarise, in the new variables ($\tilde X^i,\eta^i$) the 
tachyon vev takes the form of a non-abelian Wilson line in the NS 
sector and of a shift in position in the R-sector respectively. In 
order to obtain the generalisation of \refs{v9} to non-abelian 
tachyon vevs all we need to do is to tensor \refs{v9} with an 
element of the $u(N)$ Lie algebra. For example in the NS-sector we 
have 
\begin{equation}
        V^{(0)ia}_{T}=i\pa_{||}\tilde X^{i}\otimes\sigma_{1}
        \otimes\Gamma^{5}\Gamma^{i}\otimes t^{a}\ .
        \label{v11}
\end{equation}
In what follows we will omit the group indices
so that, for example, $\chi^i$ is understood to mean $\chi^{ia}\otimes t^a$.
The form \refs{v11} of the tachyon vertex operator suggests that 
switching on a vev for the $\chi^{i}$'s corresponds to an exact 
(in $\al'$) marginal deformation at the critical radius, provided 
$[V^{i},V^{j}]\es0$, an assertion we shall verify explicitly in the 
next section. On the other hand, in the field theory, the condition 
$[V^{i},V^{j}]\es 0$ is equivalent to $\{\chi^{i},\chi^{j}\}=0$ using 
$\{\Gamma^{i},\Gamma^{j}\}\es 0$ for $i\ne j$. But this is precisely the 
conclusion reached in \cite{LS} based on the string S-matrix 
computation. The present result shows that these non-abelian flat 
directions persist to all orders in $\al'$. We note also that it
isn't necessary that all of the radii are critical.  In order to 
perform the above change of variables we only need one radii to be
critical which then leads to a flat direction. Of course we only obtain
non-abelian flat directions if at least two of the radii are critical.

\section{Non-Abelian BI Action for Stable Non-BPS Branes}

Given that the tachyon vertex operators can be mapped into the form 
of non-abelian Wilson lines after an appropriate field 
redefinition, we can then compute the effective action within the 
existing formalism for Wilson lines 
\cite{Fradkin,BSPT,AT,Tseytlin,Callan,Perry}. The analysis is, maybe, most 
easily 
presented after T-dualising in the $4,5$-directions. In this case
the positions of the branes are also given by Wilson lines. After we
obtain the effective action for this case we can then T-dualise back 
to obtain  the effective action for
various non-BPS Dp-Branes. Now the positions are represented by the
vertex operators
\begin{equation}
V^{(0)Ia} = i\pa_{||}X^I\otimes I\otimes I\otimes t^a\ .
\end{equation}
Thus, we now 
consider $\tilde D9$-branes wrapped around $T^{4}/\bbbz_{2}\times 
T^{2}$, in the presence of a Wilson line $\A_{m}$, $m=0,1,2,...,9$,
\begin{equation}
        \A_{m}=\cases{A^{a}_{\mu}\otimes I\otimes I\otimes 
        t^{a};\quad m=0,\cdots,3\cr
        \phi^{Ia}\otimes I\otimes I\otimes t^{a}
        ;\quad m=4,5\cr
        \chi^{ia}\otimes \sigma^{1}\otimes i\Gamma^5\Gamma^{i}
        \otimes t^{a}
        ;\quad m=6,\cdots,9}
        \label{BI1}
\end{equation}
where the hermitian matrices $t^{a}$ represent the Lie algebra $u(N)$. 
Here the additional factor of $i$ in definition of 
$\chi^i$ has been inserted to ensure that all the generators in \refs{BI1}
are hermitian. 

As we have seen in the last section, at the critical radius, the 
($X^{m},\psi^{m}$) world-sheet CFT has an equivalent description in 
terms of the world-sheet fields\footnote{As only the NS sector contributes to 
the tree-level effective action we can ignore the modified boundary condition 
in the R sector in this section.}
\begin{equation}
        X^{\mu}_{R/L},X^{I}_{R/L},\tilde X^{i}_{R/L}\mtx{and} 
        \psi^{\mu}_{R/L}, \psi^{I}_{R/L},\eta ^{i}_{R/L}\ .
        \label{BI1.1}
\end{equation}
The tree level effective action is then given by \cite{AT} 
\begin{equation}
  \label{BI2}
  \Gamma(\A_m)=\<Tr P\exp\left[i\int_0^{2\pi}ds\left(\dot \Y^m
\A_m-\ha\la^m\la^n\F_{mn}\right)\right]\>\ ,
\end{equation}
with $\Y^m\in \{X^\mu,X^I,\tilde X^i\}$, $\la^m\in\{\psi^\mu,\psi^I,\eta^i\}$
and $\F_{mn} = \partial_m \A_n -\partial_n \A_m - ig[\A_m,\A_n]$. 
To continue we expand the integral in \refs{BI2} as 
($\Y^m\es Y^m+\pi^m,\; \pi^m(2\pi)=\pi^m(0)=0$)
\begin{equation}
  \label{BI3}
  \int ds \dot\xi^m A_m(Y+\pi)=\int ds\dot\pi^m[\ha\pi^n \F_{mn}+
  \frac{1}{3}\pi^n\pi^pD_p\F_{mn}+\cdots]\ ,
\end{equation}
and perform a path integration over the fields $\pi^m,\lambda^m$. Thus
the effective action for a non-BPS D-brane at the critical radius is
of exactly the same form as for a non-abelian BPS D-brane but with the
modified gauge connection \refs{BI1}. We note that, even for 
the case of a single non-BPS D-brane, where there are no $t^a$ generators,
the connection \refs{BI1} is non-abelian.  

In general, for constant, but non-commuting Wilson lines the tree-level 
effective potential will receive 
corrections to all orders in $\al'$ and we were unable to find an analytic 
expression for that case. A simplification occurs if we neglect terms of the 
form $D_{(m}F_{n)p}$ \cite{Tseytlin}. In this case the 1-loop result 
(of the $\pi^m,\lambda^m$ integration) is exact, 
leading to \cite{Tseytlin}
\begin{equation}
  \label{BI5}
  \Gamma(\A_m)=c_0\hbox{STr}\sqrt{\det(\de_{mn}+2\pi\F_{mn})}\ .
\end{equation}
Here $\hbox{STr}(M_1\cdots M_n) \es \frac{1}{n!}\sum_\sigma
\Tr(M_{\sigma(1)}\cdots M_{\sigma(n)})$ is the symmeterised trace
and $\sigma$ is a permutation. 
Note that the  determinate here is taken over $m,n$ indices and the
trace is over the Chan-Paton and cocycle factors in \refs{BI1}. 
Furthermore this expression is also valid when the 
Wilson lines are dynamical and therefore includes the kinetic terms
for the bosonic fields.

While it might be thought that the approximation 
$D_{(m}\F_{n)p}\simeq 0$ effectively assumes the background to be abelian 
it has been argued \cite{Perry} that the result may be 
valid in more general backgrounds. On the other hand, at the lowest
non-trivial order  \refs{BI5} is exact and  gives
\begin{eqnarray}
 \label{BI7}
\Gamma &=& c_0' \Tr \left(
\frac{1}{4}F_{\mu\nu}F^{\mu\nu} +\frac{1}{2}\sum_I D_\mu\phi^ID^\mu\phi^I
+\frac{1}{2}\sum_i D_\mu\chi^iD^\mu\chi^i
-V \right)\ ,\n\\
V&=&\frac{g^2}{4}\sum_{I,J}([\phi^I,\phi^J])^2
+\frac{g^2}{2}\sum_{I,j}([\phi^I,\chi^j])^2
-\frac{g^2}{4}\sum_{i\ne j}(\{\chi^i,\chi^j\})^2\ ,
\end{eqnarray}
where $c_0'$ is a constant and $D_\mu = \partial_\mu - ig[A_\mu,\ ]$.
Here we have dropped a constant 
term in $V$ and the trace is now only over
the group indices $a,b$. This precisely reproduces the bosonic part of
field theory approximation to the effective action
found in \cite{LS}  from string S-matrix calculations.

We can T-dualise the Born-Infeld action \refs{BI5} in the $x^4,\cdots,x^9$- 
directions to obtain the BI-action for a stack of stable 
non-BPS $\tilde D3$-branes on $T^4/\bbbz_2(-1)^{F_L}$ \cite {Sen,GS,LS}. 
In particular for $\phi^I\es A_\mu\es 0$ and constant $\chi^i$
\begin{equation}
  \label{BI6}
    \Gamma[\chi^i]=c_0\hbox{STr}\sqrt{\det(\de_{ij}I-2\pi i g(\Gamma^{ij})
\{\chi^i,\chi^j\})}\ .
\end{equation}
The effective potential \refs{BI6}, which is the main result of this section, 
generalises the $O(\chi^4)$-potential found in \cite{LS}. 

For constant gauge potentials $\F_{AB}=-ig[\A_A,\A_B]$. This implies 
in particular that
\begin{equation}
  \label{BI4}
  \Gamma(\A_A)=\Gamma(0,0,0)\mtx{for} [\A_A,\A_B]=0\ .
\end{equation}
Thus the the condition for marginality at tree level but to all orders in 
$\alpha'$ is simply a
zero-curvature constraint on the Wilson lines. In terms of the
scalar fields \refs{BI4} corresponds to
\begin{equation}
\label{BI8}
[\phi^I,\phi^J]=[\phi^I,\chi^j]=\{\chi^i,\chi^j\}=0\ ,
\end{equation}
for all $I,J,i,j$ with $i\ne j$.  Note that the last condition is precisely
the definition of marginality given above. In other words the 
tree level vacua of the field theory given by the zeros of the potential 
\refs{BI7} and studied in \cite{LS} are exact to all orders in $\alpha'$.

\section{The BPS States of Non-BPS Branes}

This section is devoted to some initial observations
about the BPS states on non-BPS-branes viewed as solitons  of the 
tree level effective action \refs{BI5}.  
Even though the low energy effective theory \refs{BI5} 
has no supersymmetry one may still construct
BPS bounds for the mass of soliton states in terms of conserved charges. 
While stability is not ensured by supersymmetry, we nevertheless 
expect them to enjoy some degree of stability given that they
are the lightest states carrying a given charge or satisfying a 
particular boundary condition. 

Ideally we would like to derive BPS bounds 
for the full non-linear effective 
action. However, a closed expression for the non-abelian effective
action is not known. For supersymmetric D-branes BPS bounds may be 
obtained from
the supersymmetry algebra so that the detailed form of the non-linear 
effective action is not required. However, this option is not available 
to us here. Therefore it is not clear to us how to construct a
BPS-type bound. Soliton solutions of the BPS
non-abelian Born-Infeld action have been found  in 
\cite{Brecher,Myers,Neil,DST,Zamaklar} and one might
hope that these studies can be extended to the action \refs{BI5}. 
In this section we
will therefore content ourselves by studying soliton solutions which 
saturate a BPS bound 
obtained from the lowest order term \refs{BI7} in the effective action. 
We nevertheless expect that all the states we describe below will exist 
in the full non-linear theory.

Solitons on a D-brane generally have an interpretation as intersecting
D-branes and we expect the same to be true for non-BPS branes. Therefore
before proceeding it is worthwhile to consider the stability of 
intersecting non-BPS branes. In the case of BPS D-branes it is well known
that if the open strings which stretch between two D-branes have four
coordinates with ND boundary conditions then some supersymmetry is preserved
and the configuration is stable.
For example this includes two Dp-branes intersecting over a $(p-2)$-brane
or a D$(p-2)$-brane ending on a Dp-brane. For non-BPS branes we can easily
see that the same configurations are stable  since 
open strings with ND boundary
conditions along four directions have a vanishing intercept and therefore 
no tachyons. Clearly tachyonic modes will not then be introduced
by any GSO or orbifold projection.

The first class of BPS states that may be obtained correspond to setting
$\chi^i=0$. In this case the bosonic content of \refs{BI7} is precisely
the same as for $N=2$ super-Yang-Mills theory. In particular there are three
basic types of solitons and which we will now review and re-interpret. 

In the case of a single stable non-BPS brane 
one may consider states with only the scalars $\phi^4$ and $\phi^5$ 
active. For string solitons that lie in the $x^0,x^1$ plane 
we write the energy density in the $x^2,x^3$ plane as
\begin{eqnarray}
E &=& \frac{1}{2}\int d^2x (\partial_2\phi^4)^2+(\partial_2\phi^5)^2
+(\partial_3\phi^4)^2+(\partial_3\phi^5)^2\ \n\\
&=&\frac{1}{2}\int d^2x (\partial_2\phi^4-\partial_3\phi^5)^2 + 
(\partial_3\phi^4+\partial_2\phi^5)^2 + T\ ,
\end{eqnarray}
where $T=\partial_2(\phi^4\partial_3\phi^5)
+\partial_3(-\phi^4\partial_2\phi^5)$ is a topological term. Thus the
energy is minimised for a given boundary condition if 
$\partial_2\phi^4=\partial_3\phi^5$ and $\partial_3\phi^4=-\partial_2\phi^5$. 
These 
are just the Cauchy-Riemann equations for $\phi^4+i\phi^5$ to be a
holomorphic function of $x^2+ix^3$. In the case of BPS D-branes this string
soliton corresponds to the to the intersection of two D-threebranes \cite{HLW}.
Similarly, the soliton described here corresponds to the intersection of 
two non-BPS D-threebranes whose worldvolumes lie in the $x^0,x^1,x^2,x^3$
and $x^0,x^1,x^4,x^5$ planes respectively. Although the topological term
generally gives a divergent energy contribution, there are solitons which
are smooth at their core  so that the effective action provides a reasonable 
approximation.

The other  solitons that we will discuss correspond to states which are
charged under some $U(1)$ factor of a non-abelian gauge group $U(N)$ of a 
stack of stable non-BPS branes at a critical orbifold. 
The central example
being magnetic monopoles. Here we choose the gauge $A_0=0$ and 
consider static configurations with one $\phi^I$ active. We may 
therefore write
\begin{eqnarray}
E&=& \frac{1}{2}\int d^3x \frac{1}{2}|F_{\alpha\beta}|^2\ 
+ |D_\alpha\phi^I|^2\ \n\\
&=&\frac{1}{2}\int d^3x |B_\alpha -D_\alpha\phi^I|^2 + T\ ,
\end{eqnarray}
where $\alpha,\beta=1,2,3$, $B_\alpha 
= \frac{1}{2}\epsilon_{\alpha\beta\gamma}F^{\beta\gamma}$ is the magnetic 
field and 
$T = 2\partial_\alpha(B^\alpha\phi^I)$ is a topological term. Therefore, for a 
fixed  charge given by the integral of the topological term, monopoles
satisfying $B_\alpha = D_\alpha\phi^I$ are absolute minima of the effective 
action.
For a BPS D-threebrane of
type IIB string theory the monopole corresponds to a D-string stretched
between two of the D-threebranes along $x^I$ \cite{tseytlin96,GG}. However,
since the non-BPS D-threebranes occur in type IIA string theory, 
these monopoles
correspond to non-BPS D-strings stretched between two non-BPS
D-threebranes.

There will also be electrically charged states, although these will
not have finite energy to due divergences at their core. To obtain a bound for
them we again consider static configurations with only one $\phi^I$ active
but now with $A_a=0$. In this
case the energy can be written as (using the Gauss constraint $D_0\phi^I=0$)
\begin{eqnarray}
E &=& \frac{1}{2}\int d^3x |\partial_aA_0|^2 + |\partial_a\phi^I|^2
\ \n\\
&=& \frac{1}{2}\int d^3x |\partial_a (A_0  - \phi^I)|^2 + T\ ,
\end{eqnarray}
where $T= 2\partial_aA_0\partial^a\phi^I$. We therefore find stable 
electrically charged states if $A_0=\phi^I$ and 
$\partial_a\partial^a \phi^I=0$. These simply correspond to fundamental
strings which end on the non-BPS D-threebrane an are infinitely
extended along $x^I$ \cite{CM,HLW2,Gibbons}. 

Let us now consider solitons involving the scalar fields $\chi^i$ instead
of the scalars $\phi^I$. 
The string soliton mentioned above no longer exists due to the presence
of the $(\chi^i\chi^j)^2$  term in the potential of \refs{BI7}. However
the charged solutions certainly do exist since if only one $\chi^i$ is
active then the potential term vanishes just as it did above for $\phi^I$. 
Thus there are BPS saturated monopole states with $B_a = D_a\chi^i$ 
and electric states with $A_0=\chi^i$ too.
 
Note that turning on  the tachyon
vev $\chi^i=1$ corresponds to deforming the non-BPS D-threebrane into a 
D-fourbrane/anti-D-fourbrane pair 
wrapped along $x^i$ with half a unit of Wilson line \cite{sen9904207}. 
Therefore a monopole soliton
involving $\chi^i$ can be pictured as two 
non-BPS D-threebranes at the core which at infinity fatten out into two 
D-fourbrane/anti-D-fourbrane pairs wrapped around $x^i$, with a 
non-abelian Wilson line turned on.
These states again correspond to non-BPS D-strings but this time
stretched along an orbifold direction $x^i$. 
Similarly the most natural interpretation for the 
electric states is that 
they correspond to a fundamental string stretched along
$x^i$ and ending on the non-BPS D-threebranes. 

Next we consider what happens if we move away from the critical radius. 
In this case the
mass term $\frac{1}{2}m_i^2(\chi^i)^2$ appears in the action \refs{BI7}
\cite{LS}. The solutions we just described involving $\chi^i$ then
no longer exist. In the case of the monopole 
this can be understood because a non-BPS
D-threebrane is stable and hence 
the $m_i^2$ are positive only if the orbifold radii are
larger than the critical radius. However a non-BPS D-string stretched along
an orbifold direction is stable only if the radius is less than critical.
Therefore the two can only be simultaneously stable precisely at the
critical radius. For the electrically charged states, 
however, it is not clear to us what the origin of the instability is. 

Finally we note that   
Yang-Mills theories possesses other solitons including the
so-called $1/4$-BPS states with two scalars 
active \cite{quarterBPS}. 
Clearly these solitons also appear in the action \refs{BI7}. 
In general though these solitons won't exist if
both scalars are taken to be tachyon modes. In addition to the states
we described above  
there will also be ``mixed'' 
solitons where some of the $\phi^I$ and only one of the $\chi^i$
scalars  are active since in these cases the effective action is same as  
$N=4$ super-Yang-Mills. 
We would also like to  highlight the possibility that 
the effective action \refs{BI7} 
admits new types of  solitons 
associated with the non-abelian flat directions in the tachyon potential.

\section*{Acknowledgements}

The authors wish to thank G. Watts for helpful discussions and 
collaboration in the early stages of this paper. I.S would like to thank 
King's College London, for hospitality during the initial stages of this work. 
N.D.L. is supported by a PPARC Advanced Fellowship and also in 
part by the PPARC grant PPA/G/S/1998/00613.
He would like to thank the Theory Division at CERN for its hospitality
while this work was completed.

\end{document}

\end